\documentclass[preprint,aps,showpacs]{revtex4}
\begin{document}
\title{Double heavy baryons and dimesons}
\author{Mitja Rosina}
\affiliation{Faculty of Mathematics and Physics, University of Ljubljana,\\
     Jadranska 19, P.O.~Box 2964, SI-1001 Ljubljana, Slovenia,\\
and Jo\v{z}ef~Stefan Institute, Ljubljana, Slovenia}
\author{Damijan Janc}
\affiliation{Jo\v{z}ef~Stefan Institute, 
              Ljubljana, Slovenia\vskip 2cm}
\begin{abstract}
 {\begin{centering}
{ We critically examine the question whether
 \\
the
cc$\bar{{\rm q}}\bar{{\rm q}}$ dimeson is bound or not.\\\vskip 9cm}  
\end{centering}\vskip 1cm }
\end{abstract}
\maketitle
%%%%%%%%%%%%%%%%%%%%%%%%%%%%%%%%%%%%%%%%%%%%
%% MAINMATTER
%%%%%%%%%%%%%%%%%%%%%%%%%%%%%%%%%%%%%%%%%%%%

\section{Introduction}
There is a revived interest to study double heavy baryons and dimesons,
both due to the theoretical urge of understanding better the quark-quark
effective interaction, as well as due to new experimental opportunities
in Fermilab and LHC.

The effective interaction between heavy quarks (and antiquarks) is 
expected to be cleaner than between light quarks. For heavy particles 
the nonrelativistic constituent quark model is more acceptable, 
the perturbative QCD contributions (such as one-gluon-exchange) 
are more adequate and chiral fields are less important.
The effective interaction between a heavy quark and a heavy antiquark
has been reasonably well studied and fitted by the charmonium and bottomium
spectra. There is, however, no free diquark to study the effective interaction
between two heavy quarks; one has to dress the diquark in order to
obtain a color singlet object QQq or QQ$\bar{{\rm q}}\bar{{\rm q}}$
(Q=c or b, q=u,d, or s).

It is straightforward to extrapolate the one-gluon-ex\-change (OGE) 
interaction from Q$\bar{{\rm Q}}$ to QQ (Q= any quark). 
The charge conjugation
changes the $\bar{{\rm Q}}$ antitriplet to Q triplet. Then the color
factor $\lambda\cdot\lambda/4 = -4/3$ for the Q$\bar{{\rm Q}}$
singlet changes to $-2/3$ for the QQ antitriplet 
(the ``$V_{{\rm QQ}}={1\over2}V_{{\rm Q}\bar{{\rm Q}}}$ {\it rule''}).
On the other hand, it is questionable whether the (linear) confining
potential should also possess such a color factor and obey the
$V_{{\rm QQ}}={1\over2}V_{{\rm Q}\bar{{\rm Q}}}$ rule. The fact that
the ground state energies and some excited states of light and heavy
baryons are reasonably well reproduced with such a ``universal''
OGE + confining effective interaction is encouraging \cite{SB1}
but not conclusive. There may be other mechanisms for the 
$V_{{\rm QQ}}={1\over2}V_{{\rm Q}\bar{{\rm Q}}}$ rule. For example,
the flux tubes in a Y configuration can be mimicked by
twice weaker two-body flux lines since the length of the arms of the
Y is approximately half the length of the circumference
of the triangle. The color singlet 3-quark system is insensitive to the 
features of the $colour\cdot colour$ operator since it is just 
a constant in the 3-body singlet representation. To explore the
color structure of the effective interaction one has to go beyond
mesons and baryons to dimesons and other exotics.

The study of double-heavy baryons and of double-heavy dimesons are
complementary. The double-heavy ba\-ryons help to study
the QQ interaction, while the dimesons also test the pion exchange 
between light quarks \cite{GRAZ} and are more sensitive to
three-body forces.

Our constituent quark model calculation \cite{Janc} has shown
the bb-dimeson to be bound by more than 100 MeV and the cc-dimeson
to be unbound, which is consistent with some other calculations,
for example \cite{SB}. We have proposed to look for the bb-dimesons
at LHC assuming a mechanism of double b$\bar{\rm b}$
production by double gluon gluon fusion
$\rm (g+g)+(g+g)\to(b+\bar{b})+(b+\bar{b})$ which has been described
at this Conference by Danielle Treleani \cite{Treleani}.
The two b-quarks then join into a diquark which gets dressed with 
a light quark or two light antiquarks to become a double heavy baryon 
or a dimeson. However, the production rate bb-dimesons has been
estimated to be rather low \cite{LHC1,LHC2}, about 5 events/hour,
and there seem to be no characteristic decays.

Therefore it is of utmost importance to look also for the cc-dimesons
since their production rate might be as much as 10$^4$ events/hour if 
the same mechanism applies. They would also be easier to detect, 
for example by  $\rm cc\bar{u}\bar{d}\to D^+ + K^- +\pi^+$.
There is, of course, a grat risk that they do not exist. If they,
however, do exist they would be very exciting --
we would have to revise our ideas about the effective quark-quark 
interaction, and/or introduce many-quark forces.

\section{Can the ${\rm cc\bar{u}\bar{d}}$ dimeson be bound?}

We have obtained a phenomenological estimate for the binding
energy of the cc-dimeson ($ISP=01+$) with respect to the DD$^*$ 
by assuming a compact structure like in the $\bar{\Lambda}_{\rm c}$
or $\bar{\Lambda}_{\rm b}$ baryon with the cc-diquark playing
the role of the heavy antiquark. For the cc binding in the
diquark we assumed the
$V_{{\rm QQ}}={1\over2}V_{{\rm Q}\bar{{\rm Q}}}$  rule and no 3-body forces.
We compared the following hadrons \cite{Janc}

\begin{eqnarray*}
m_{{\rm cc}\bar{{\rm u}}\bar{{\rm d}}} &=& 
   2m_{{\rm c}} + m_{{\rm u}} +m_{{\rm d}} + E_{cc} 
  + E_{\bar{{\rm u}}\bar{{\rm d}} [{\rm cc}]} \\
m_{{\rm J}/\psi}  &=& 2m_{{\rm c}}  + E_{{\rm c}\bar{{\rm c}}} \\
m_{\bar{\Lambda}_{{\rm c}}} &=& m_{{\rm c}} + m_{{\rm u}} +m_{{\rm d}}
  + E_{\rm \bar{u}\bar{d}\bar{c}}
\end{eqnarray*}
where $E_{\bar{{\rm u}}\bar{{\rm d}} [{\rm cc}]}\approx 
E_{\rm \bar{u}\bar{d}\bar{c}}$ is the
potential plus kinetic energy contribution of the two light antiquarks 
in the field of a heavy diquark or antiquark, respectively,  
and it cancels in the difference in the limit where the mass 
of the b quark goes to infinity and the heavy diquark is point-like
so that we can neglect the size of the heavy diquark in the dimeson.

We estimated the diquark binding energy by using the theorem \cite{Janc}
$V_{{\rm cc}}= \frac{1}{2} V_{{\rm c\bar{c}}} \quad \Rightarrow \quad
E_{{\rm cc}}(m_{{\rm red}})
= \frac{1}{2} E_{{\rm c\bar{c}}}(\frac{1}{2} m_{{\rm red}})$.
Since meson binding energies lie on a smooth curve
as a function of their reduced masses, it is easy to interpolate
for the "fictitious meson" with $m_{{\rm red}}/2$ and we get \cite{Janc}
\newline
$E_{{\rm cc}}-\frac{1}{2} E_{{\rm c\bar{c}}}=134\pm20\,{\rm MeV}$ yielding
\begin{eqnarray*}
\Delta E_{{\rm cc}\bar{{\rm u}}\bar{{\rm d}}}
   &=&m_{\Lambda_{{\rm c}}}+m_{J/\psi}/2 +
   E_{{\rm cc}}-E_{{\rm c\bar{c}}}/2
   -m_{{\rm D}}-m_{{\rm D^*}}\\
   &=& (-42+134)\, {\rm MeV} = +92\, {\rm MeV}.
\end{eqnarray*}
This means that such a compact structure is not bound with respect 
to the DD$^*$ threshold. Also detailed four-body calculations 
with OGE+linear potential with 
Bhaduri or Grenoble parameters \cite{SB} did not yield a bound state.

An alternative estimate lies considerably lower but is still unbound:
\begin{eqnarray*}
\Delta E_{{\rm cc}\bar{{\rm u}}\bar{{\rm d}}}
   &=&m_{\Lambda_{{\rm b}}}-m_{\rm b}+m_{\rm c}+m_{J/\psi}/2 +
   E_{{\rm cc}}-E_{{\rm c\bar{c}}}/2\\
   &-&m_{{\rm D}}-m_{{\rm D^*}}
   = (-94+134)\, {\rm MeV} = +40\, {\rm MeV}.
\end{eqnarray*}
The actual cc-diquark mass lies midway between the masses of
the c and b quark (appearing in the center of $\Lambda_{\rm c}$
and $\Lambda_{\rm b}$, respectively), therefore the answer is 
inbetween the two estimates which still means no binding.

The question arises whether the parameters in the OGE+linear confinement
model could be stretched so as to bind cc-dimeson 
without spoiling the fit to mesons and baryons.
If the $V_{{\rm QQ}}={1\over2}V_{{\rm Q}\bar{{\rm Q}}}$ rule applies
smaller quark masses could do the job. 
For Bhaduri masses, half of reduced mass od the cc diquark 
($m_{\rm c}/4=467\,{\rm MeV}$) coincides with the reduced mass of 
${\rm  D_s}$,
$m_{\rm c} m_{\rm s}/(m_{\rm c}+m_{\rm s})=454\,{\rm MeV}$ so that
$E_{\rm cc}= \frac{1}{2} E_{\rm c\bar{s}}$. If we decrease all quark masses
by 200 MeV, the reduced mass of ${\rm  D_s}$, would decrease by 132 MeV and
$m_{\rm c}/4$ only by 50 MeV. Higher reduced mass of cc 
compared to ${\rm  D_s}$ means better binding of cc (by about 40 MeV).
This is still not quite enough but might work in cooperation
with additional effects.

A three-body interaction of the type
\begin{equation}
V_{ijk}=-\frac{U_0}{8} d^{abc}\lambda_i^a\lambda_j^b\lambda_k^c\,
        \exp(-(r_i^2+r_j^2+r_k^2)/a^2)
\end{equation}
with at most $U_0=20$ MeV and $a=2.3$ fm would bind. 
The choice of $a<1\,{\rm fm}$ gives  small effect, and above 2.3~fm
the effect saturates. Due to the combinatorics,
a three-body interaction is more effective for tetraquarks 
than for baryons and the proposed one spoils baryons only by few MeV.

The pion exchange between D and D$^*$ leads to a coulomb-like 
long-range force because the exchanged pion is almost on the mass shell 
\cite{Richard}:
$({\rm D}^*\to {\rm D}+\pi),\;( {\rm D}+\pi \to {\rm D}^*)$.
(Note that $m_{\rm D^{*+}}-m_{\rm D^+}-m_{\pi^0}= 5.6\,{\rm MeV},\quad
m_{\rm D^{*0}}-m_{\rm D^0}-m_{\pi^0}=7.1\,{\rm MeV},\quad
m_{\rm D^{*+}}-m_{\rm D^0}-m_{\pi^+}= 5.8\,{\rm MeV}.)$
This should in principle give a (weak) binding.
We are studying the conflicting effects of short-range QQ interaction
and this long-range ${\rm DD}^*$ interaction.

\section{A speculation using the ccu and ccd signals}

Recent SELEX experiments and analyses \cite{SELEX} gave some
more and some less convincing signals about the ccu(3460 and
3541) and ccd(3443 and 3520) baryons.
If confirmed, they would have a dramatic effect on our estimates
about the binding of the ${\rm cc\bar{u}\bar{d}}$ dimeson.
If refuted, the present section remains  a piece of science fiction.

Our expectations about the ccq baryon are consistent with the 
$\sim 3530$ MeV isodoublet but would need a lot of stretching 
to accommodate the $\sim 3450$ isodoublet 
(if this one is confirmed as the spin=1/2 ground state).
A phenomenological estimate similar as in the previous section gives
for s=1/2 (assuming an S=1 cc-diquark) the value inbetween
$$m_{\rm ccq}=\frac{1}{2}m_{{\rm J}/\psi}
         +E_{\rm cc}-\frac{1}{2}E_{\rm c\bar{c}}
         +\frac{3}{4}m_{\rm D}+\frac{1}{4}m_{\rm D^*}=3584\,{\rm MeV}$$
and
\begin{eqnarray*}
m_{\rm ccq}&=&\frac{1}{2}m_{{\rm J}/\psi}
         +E_{\rm cc}-\frac{1}{2}E_{\rm c\bar{c}}+m_{\rm c}-m_{\rm b}\\
         &+&\frac{1}{4}m_{\rm B}+\frac{3}{4}m_{\rm B^*}
         -\frac{1}{2}(m_{\rm D^*}-m_{\rm D}) = 3535\,{\rm MeV}
\end{eqnarray*}

The predicted spin 3/2 state lies higher by \newline
$\frac{3}{4}(m_{\rm D^*}-m_{\rm D})=106$ MeV
Such spin-spin splitting is noticeably larger
than the difference 80 MeV between the 3530 and 3450~MeV SELEX levels  
and it will be some surprise if the 3450 level is confirmed 
as a ground state and the 3530 level gets an 3/2 assignment.

Then follows a phenomenological estimate for the cc-dimeson
\begin{eqnarray*}
\Delta E_{{\rm cc}\bar{{\rm u}}\bar{{\rm d}}}
   &=& m_{\rm ccu}-(\frac{3}{4}m_{\rm D}+\frac{1}{4}m_{\rm D^*})\\
   &+&m_{\Lambda_{{\rm c}}} -m_{{\rm D}}-m_{{\rm D^*}}\\
   &=& -42 \quad{\rm or}\quad +38\, {\rm MeV} 
\end{eqnarray*}
assuming the 3450 or 3530 MeV level, respectively, 
to be the ccu ground state

The alternative estimate is very similar. \newpage
\begin{eqnarray*}
\Delta E_{{\rm cc}\bar{{\rm u}}\bar{{\rm d}}}
   &=& m_{\rm ccu}-(\frac{1}{4}m_{\rm B}+\frac{3}{4}m_{\rm B^*})\\
   &+& \frac{1}{2}(m_{\rm D^*}-m_{\rm D})
    + m_{\Lambda_{{\rm b}}} -m_{{\rm D}}-m_{{\rm D^*}}\\
   &=& -45 \quad{\rm or}\quad +35\, {\rm MeV} 
\end{eqnarray*}

\section{Conclusion}

There are several subtle effects each of which separately 
is not likely to bind the ${\rm cc\bar{u}\bar{d}}$ dimeson
with respect to the DD$^*$ threshold.
However, their cooperative effect might just bind it or just
fail to bind it. Therefore we join and support those researchers
who propose the detection of the 
${\rm cc\bar{u}\bar{d}}$ dimeson as a crucial experiment.

\end{document}